\documentclass[superscriptaddress,aps,prl,floats,twoside,floatfix]{revtex4}
\usepackage{epsfig}
\usepackage{amsmath}
\usepackage{amssymb}
\usepackage[english]{babel}

\usepackage{amsfonts,amssymb,graphicx}
\usepackage{epstopdf}

\usepackage{indentfirst}

\usepackage{amsthm}
\usepackage{amsmath}
\usepackage{latexsym}
\usepackage{amsthm}

\usepackage{amsmath, amssymb, mathrsfs, dsfont, mathtools, empheq,bm,bbm}
\usepackage{color}

\def\be{\begin{equation}}
\def\ee{\end{equation}}

\newcommand{\PP}{\mathbb{P}}
\newcommand{\R}{\mathbb{R}}
\newcommand{\E}{\mathbb{E}}
\newcommand{\N}{\mathbb{N}}
\newcommand{\balpha}{\bm{\alpha}}
\newcommand{\bbeta}{\bm{\beta}}

\begin{document}

\title{Equilibrium and Dynamics of a Multi-Bath Sherrington-Kirkpatrick Model}
\author{Pierluigi Contucci}
\affiliation{Universit\`{a} di Bologna, Piazza di Porta S.Donato 5, 40127 Bologna, Italy}
\author{Jorge Kurchan}
\affiliation{ Laboratoire de Physique de l'Ecole Normale Sup\'erieure, ENS, Universit\'e PSL, CNRS, Sorbonne Universit\'e, Universit\'e Paris-Diderot, Sorbonne Paris Cit\'e, Paris, France}
\author{Emanuele Mingione}
\affiliation{Universit\`{a} di Bologna, Piazza di Porta S.Donato 5, 40127 Bologna, Italy}

\begin{abstract}
In this paper we study the equilibrium statistical mechanical  as well as the dynamical properties of a Sherrington and Kirkpatrick model in a multi-bath setting introduced in \cite{cuku}. We show that the free energy per particle in the thermodynamical limit obeys a variational principle of Parisi type. The relation between the resulting order parameters is discussed.
\end{abstract}


%

\maketitle
\vskip 1 cm
\rightline{\textit{To Giorgio Parisi on his 70-th birthday}}

\section{Introduction}
The equilibrium statistical mechanics of a disordered system (in general a system where the  degrees of freedom can be divided in two different families) is described with  two extreme prescriptions known, in the literature, as \textit{quenched} and \textit{annealed}. The spin-glass case, for instance, is defined by the quenched measure where the random coupling disorder $J$ is kept fixed while the spins $\sigma$ are thermalised according to the Boltzmann distribution. This perspective is considered physically relevant because the relaxation time of the disorder interaction variables is much slower than the one for the spin variables. Conversely, in the annealed prescription the disorder variables thermalise together with the spin ones. However one can also consider a different equilibrium measure depending on a real positive number $\zeta$ with thermodynamic pressure
\be\label{tala}
P(\zeta) \; = \; \frac{1}{\zeta}\log \E Z_J^\zeta \; ,
\ee
where $Z_J=\sum_{\sigma}e^{-\beta H(\sigma)}$ is the partition function given $J$, a random variable depending on the disorder $J$ obtained integrating on the spins. By definition we have that $P(0)$ corresponds to the quenched equilibrium while $P(1)$ to the annealed one. Hence  $\zeta$ can be viewed as a {\it scale} parameter in the unit interval interpolating between the quenched case with annealed one \cite{BGM}.
The origins of \eqref{tala}  are to be found on the replica approach to spin glasses \cite{MPV} where $\zeta$ is at the outset an integer. Almost thirty years ago Kondor \cite{kondor} calculated \eqref{tala} using the replica trick for the partition function of  the Sherrington Kirkpatrick model. He found that the computation is essentially the usual one for a Parisi ansatz $x(q)$, but with $\zeta \leq x\leq 1$. This does not mean that the solution itself is the usual $x(q)$ truncated at $x = \zeta$,  but rather that the Parisi equations have to be solved in this interval. A rigorous proof of this result has obtained by Talagrand  in  \cite{tal}.

In this work we consider  a generalization of  \eqref{tala} consisting  a multi-scale equilibrium measure obtained by successive independent integration on a suitable class of variables. The idea
of studying a system at different energy scales is common in physics at least since the early days of the Euclidean approach to renormalisation group in quantum field theory (see \cite{Pol,Gal}). For two scales
$\zeta_0$ and $\zeta_1$ the model is defined in terms of an interaction $J=(J_0,J_1)$ with independent components:
\be
e^{\zeta_1 P^{(0)}} \; = \;  \E_1 Z_J^{\zeta_1} \; ,
\ee
and
\be
e^{\zeta_0 P} \; = \;  \E_0\, e^{\zeta_0P^{(0)}} \; .
\ee
For $r$ scales $\zeta_0< \zeta_1<\ldots<\zeta_{r-1} < \zeta_{r}\,=\,1$
the recursion relations are
\be\label{hier}
e^{\zeta_{l} P^{(l-1)}} \; = \;  \E_l\, e^{\zeta_lP^{(l)}} \; ,
\ee
where $0 \le l \le r$, $\E_r\,e^{P^{(r)}}=Z_J$ and $P=P^{(-1)}$.

The relation \eqref{hier} is also referred to as multi-bath in \cite{cuku,cuku2} where it was applied to  a class of   out of the equilibrium dynamical systems
in the limit of small entropy production. Those papers presented analytical and numerical evidences suggesting that one can introduce several effective temperatures and partial thermalizations derived from  the functional form \eqref{hier} such that correlation and response functions behave as in a multiple equilibrium bath. In \cite{Mon} the functional form \eqref{tala} is used to get information about the landscape of the free energy of a given statistical mechanics model. More precisely, for a model with Hamiltonian $H(\sigma)$, the idea is to introduce an auxiliary field  $\phi$ weakly pinned with $\sigma$ but frozen. Keeping fixed the field $\phi$, the variables  $\sigma$ thermalize in the usual Gibbs setting at inverse temperature $\beta$. Then  the resulting free energy is used as an effective Hamiltonian function acting on the field $\phi$  at temperature $\zeta$. The examples given  in \cite{Mon} show how the previous construction can detect hidden metastable states of a glassy system.

An important observation is that under quite general assumptions on the randomness it is possible to express the  relation \eqref{hier} by coupling a suitable Hamiltonian function with a Ruelle probability cascade (RPC) \cite{Derrida,Ruelle,panbook,cmp}). This structure is the core of the Guerra's Replica Symmetry Broken bound \cite{broken} for the SK model where a random field, described by a linear covariance, is coupled to a RPC.

In this work instead we consider and exactly solve a disordered mean field model where the Hamiltonian includes an SK term ( two-spin interaction) and is coupled with a given RPC. From the physical point of view the effect of  coupling a  disordered Hamiltonian with a RPC corresponds to replacing  the standard Gibbs equilibrium measure with a \textit{multi-bath} thermalisation defined through a hierarchical sequences of averages \eqref{hier} tuned by several effective "temperatures". In section 2 we introduce the model and presents its solution. It turns out that the thermodynamical limit of the free energy can be expressed as a Parisi-like variational problem.  In section 3 we give a dynamical interpretation of the multibath equilibrium and discuss some aspect of the solution we found.

\section{The model and the solution}

Given $N \geq 1$ let us consider a  system  of $N$ spins $\sigma=(\sigma_i)_{\,i\leq N }\in\Sigma_{N}=\{-1,1\}^N$.  Fix an integer $r\geq1$  and a two sequences

\be\label{zetaseq}
0=\zeta_{-1}<\zeta_0< \zeta_1<\ldots<\zeta_{r-1} < \zeta_{r}\,=\,1
\ee

\be\label{gammaseq}
0=\gamma_0<\gamma_1<\ldots<\gamma_{r}=1
\ee

%
%
%
%
%
%
%
%
%
%
%
%

For any $l=1\ldots,r$  consider an independent  random Hamiltonian function

\be\label{hami}
H_N(\sigma,l)\,=\, -\,\sqrt{\frac{\gamma^2_{l}-\gamma_{l-1}^2}{N}}\sum_{i,j=1}^N J^{(l)}_{ij}\sigma_i\sigma_j-h\sum_{i=1}^N \sigma_i
\ee
where $J^{(l)}_{ij}$ is family of independent  centered gaussian with variance $\gamma^2_{l}-\gamma_{l-1}^2$
and $h\in\R$ is a given external magnetic field. The family $H_N(\sigma,l)$ is a gaussian process on  $(\sigma,l)\in\Sigma_N\times \{1,\ldots,r\}$ with covariance

\be\label{coint}
\E \,H_N(\sigma^1,l) H_N(\sigma^2,l) \,=\, N\,\,\left(\sqrt{\,\gamma^2_l-\gamma^2_{l-1}\,}\, q_N(\sigma^1,\sigma^2)\right)^2\,
\ee

where
\be\label{overlap}
q_N(\sigma^1,\sigma^2)\,=\,\frac{1}{N}\sum^N_{i=1}\,\sigma^1_i\,\sigma^2_i
\ee

is the usual overlap. Then we define the pressure density as
\be\label{multiscalep2}
p_N= \frac{1}{N}\,\log Z_{0,N}
\ee

where  $Z_{0,N}$ is obtained recursively in the following  the general scheme in \eqref{hier}. We denote
by $\E_l$ denotes the average w.r.t. the randomness in $H_N(\sigma,l+1)$  and starting from

\be
Z_ {r,N}\, =\,\sum_{\sigma}\, \prod_{1\leq l\leq r}\,e^{\,-\beta H_N(\sigma,l)}
\ee

we define
\be\label{recZ}
Z^{\zeta_{l-1}}_{l-1,N}\,=\,\E_{l-1}\,Z^{\zeta_{l-1}}_{l,N}
\ee
for any $0\leq l\leq r-1$.\\

We mention that for $r=1$ and a generic $\zeta_0>0$ the model was studied and solved by Talagrand in \cite{tal}. If $\zeta_0 \to 0$ we recover the SK model at inverse temperature $\beta$.

The recursive definition \eqref{multiscalep2} of the pressure per particle  entails an equilibrium measure characterized by a hierarchical sequence of averages. As example, by   \eqref{recZ} and using gaussian integration by parts,  the internal energy of the system
can be represented as follow
\be\label{internal}
\frac{\partial p_N}{\partial \beta}\,=\frac{\beta^2}{2}\left(1-\sum_{l=1}^r (\zeta_l-\zeta_{l-1})\gamma_l^2\,\big\langle q_{12}^2\,\big\rangle_l\right)
\ee

where $q_{12}$ is the overlap \eqref{overlap} and $\langle\,\rangle_l$ is a suitable average ( see \cite{cmp, broken} for a precise definition)  representing the partial thermalization at the level $l$.

The  \textit{multi-bath} equilibrium   can  be described as follows: we start supposing that the spin variables thermalize at inverse temperature $\beta$ and then , for any $l=1,\ldots,r$, we thermalize the  random couplings $J^{(l)}$ with an  inverse temperature $\beta/\zeta_{l-1}$ and an effective Hamiltonian $\log Z_{l,N}$ an so on.

\subsection{The order parameter and the variational problem}

It turns out \cite{cmp} that the order parameter of the model is a distribution function on $[0,1]$,
however, comparing to the  Sherringhton-Kirkpatrick model, here  it  must
satisfy an additional condition that preserves the multibath structure fixed at the beginning (see eq. \eqref{internal}). Consider an arbitrary integer $k\geq r$ and a sequence $\xi=(\xi_j)_{j\leq k}$ such that
\be\label{newseq}
0=\xi_{-1}<\xi_{0} < \xi_{1}< \ldots < \xi_k\,=\,1\
\ee

Moreover we assume that
\be\label{spacecon}
\zeta\subseteq\xi
\ee

Now given the sequence $\xi$ in \eqref{newseq} consider  the following  subset of $\{0,\ldots,k\}$

\be\label{mergeset}
K_l\,=\,\left\{\,j\,:\, \zeta_{l-1}< \xi_j\leq\zeta_{l},\,0\leq j\leq k\right\}
\ee

for any $l\leq r$. Given the sequence $\gamma$ in \eqref{gammaseq} we construct a new sequence $\widetilde{\gamma}=(\widetilde{\gamma}_j)_{ j\leq k}$ defining for any  $j\leq k$

\be\label{seqq}
\widetilde{\gamma}_j\,=\,\gamma_l\,\,\,\, \mathrm{if}\,\,\, j\in K_l
\ee

We also introduce  an arbitrary sequence $q=(q_j)_{j\leq k}$ such that

\be\label{seqc}
0\,=\,q_0  \leq q_{1} \leq \ldots \leq q_{k}=1
\ee

and we define a sequence

\be\label{sec2}
c=(\tilde{\gamma_j}q_j)_{j\leq k}
\ee

Given the sequences $\xi$ and $c$  the order parameter of the model is the piecewise constant function $x: [0,1]\to[0,1]$ defined as

\be\label{order}
x(c)= \xi_j,\,\,  c\,\in [c_{j-1},c_j)
\ee

for any $j=1,\ldots, k$ and  $x(1)=1$.

Then the  order parameter $x(c)\in X_{\zeta}$,   where $X_{\zeta}$ is the space of  distribution
functions on [0,1] that contains $\zeta$ in the image. Physically  $x(c)$ represents the distribution of the overlap w.r.t. the multi-bath equilibrium measure previously introduced.  Notice  that conditioning on the event $K_l$ you get

\begin{align}
\E (c^2) = \sum_{l=1}^r\PP(K_l)\,\gamma^2_l \,\E(q^2|K_l)=\\ \nonumber
=\sum_{l=1}^r\,(\zeta_l-\zeta_{l-1})\,\gamma^2_l \,\E(q^2|K_l)\
\end{align}

that corresponds to the finite volume decomposition in \eqref{internal}.

For example in the case  $r=2$ the model is determined by the parameters $\zeta_0,\zeta_1$ and $\gamma_1$ and  a possible choice for $x(c)$  with  $k=6$ is showed figure 1. In this case we have that the sets defined in \eqref{mergeset} are

\be
K_2=\{6,5,4\},\,\,\,K_1=\{3,2,1\},\,\,K_0=\{0\}
\ee
and then \eqref{seqq} leads to

\be
\widetilde{\gamma}_{6,5,4}=1,\,\,
\widetilde{\gamma}_{3,2,1}=\gamma_1,\,\,
\widetilde{\gamma}_0=\gamma_0
\ee

We want to emphasize that two important properties of $x$ that holds for any choice of $r$, $\zeta$ and $\gamma$

- condition \eqref{seqq} implies that  $\lim_{c \to 0^{+}}x(c)=\zeta_0$, namely there is always a jump discontinuity in $0$ with gap $\zeta_0$.\\

- any possible order parameter $x$ has at least $r$ levels of replica symmetry breaking. \\

In the next section   we will give a physical interpretation  of the above properties in the dynamical framework of the multibath equilibrium  \cite{cuku,cuku2}.

\begin{figure}\label{fig1}
\includegraphics[scale=0.3]{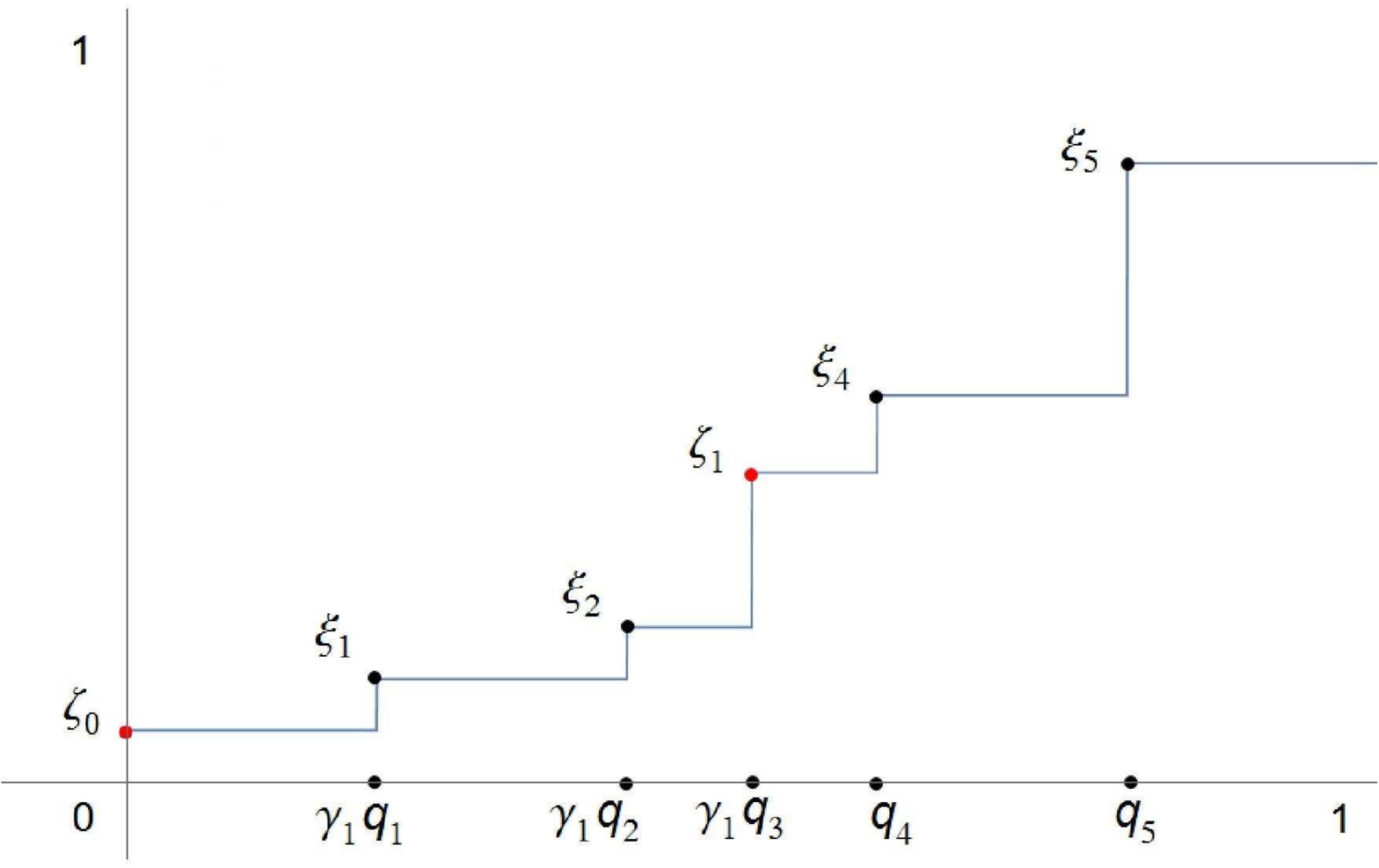}
\caption{An example of an allowed trial parameter $x(c)$ in the case r=2 and k=6, the $\zeta$ coordinates are the pinning points given a priori by the multi-bath equilibrium measure.}
\end{figure}

Finally the solution of the model is given in terms of the following  variational problem.

Let $f(c,y;x)$ be a function of the variables $c\in[0,1]$, $y\in \R$, depending also on the order parameter $x$  as the solution of the Parisi PDE
\begin{equation}
\label{antipara} \frac{\partial f}{\partial c} \,+\,
{1\over2}\left( \frac{\partial^2 f}{\partial y^2}\,+\,x(c)\left({\frac{\partial f}{\partial y}}\right)^2\right)=0
\end{equation}
with final condition
\begin{equation}
\label{final} f(1,y)=\ln\cosh(\beta y)
\end{equation}

We define the Parisi functional for the Multiscale SK model the quantity

\be\label{parisifun}
\mathcal{P}(x)\,=\,\log\,2\,+\,f(0,h;x)\,-\,\frac{\beta^2}{2}\int^1_0 x(c)c \,dc
\ee

Then it's possible to prove \cite{cmp} that the thermodynamic limit of the quenched pressure density of the Multiscale SK model $p_N$   exists and is given by
\be\label{main}
\lim_{N\to\infty}p_{N}\,= \,\inf_{x\in X_\zeta}\,\mathcal{P}(x)
\ee

\subsection{Sketch of the proof}

The strategy is to start with the  following   representation of the recursive definition \eqref{recZ}.
Let us denote by $\balpha\in\N^r$  an auxiliary  degree of freedom of the system. A configuration is now
is
\be\label{conf}
(\sigma,\balpha)\in \Sigma_N\times\N^r\,\equiv\,\Sigma_{N,r}
\ee

Let $(\nu_{\balpha})_{\balpha\in\N^{r}}$ be the random weights of Ruelle Probability Cascade associated to the sequence $\zeta$ (see \cite{cmp}). For $\balpha,\bbeta\in \N^r$ we denote

\be\label{wegge}
\balpha\wedge\bbeta\, = \min\,\{0\leq l \leq r\,|\,\alpha_1 = \beta_1,\ldots , \alpha_l = \beta_l, \alpha_{l+1} \neq \beta_{l+1}\}
\ee

where $\balpha\wedge\bbeta=r$ if $\balpha=\bbeta$. It's useful to think $\N^r $ as the set of leaves of an infinite tree $\mathcal{A}=\N^0\cup\N\cup\N^2\ldots\cup \N^r $  of depth $r$ and root
$\N^0=\{\emptyset\}$. Then  $\balpha\wedge\bbeta$ denotes  the level of their common ancestor.

Let $\Big(g(\balpha)\Big)_{\balpha\in\N^r}$ be a family of centered gaussian random variables with covariance
\be
\E \,g(\balpha^1)\,g(\balpha^2) = \,\left(\gamma_{\balpha^1\wedge\balpha^2}\right)^2\,
\ee

Consider a gaussian process $H_N$ on $\Sigma_{N,r}$ defined by

\be\label{hamrpc}
H_{N}(\sigma,\balpha)\,=\,\frac{1}{\sqrt{N}}\,\sum_{i,j=1}^N\,g_{ij}(\balpha)\,\sigma_i\sigma_j
\ee

where  $\left(g_{ij}(\balpha)\right)_{i,j=1,\ldots ,N}$ is a family of i.i.d. copies of $g(\balpha)$.\\

Given two configurations $(\sigma^1,\balpha^1),\,(\sigma^2,\balpha^2)\in\Sigma_{N,r}$  the covariance of the process $H_{N}$ is

\be\label{covH}
\E \,{H}_N(\sigma^1,\balpha^1)\,{H}_N(\sigma^2,\balpha^2) \,=\, N\,\left(c_{12}\right)^2
\ee

where
\be\label{overl}
c_{12}\,=\,{\gamma}_{\balpha^1\wedge\balpha^2}\,q_N(\sigma^1,\sigma^2)
\ee

Notice that $q_N(\sigma^1,\sigma^2)$

is  the usual  \textit{overlap} between two configurations $\sigma^1,\sigma^2\in\Sigma_N$
while ${\gamma}_{\balpha^1\wedge\balpha^2}$ is the overlap between two points in the space $\N^r$ equipped with the random measure $(\nu_{\balpha})_{\balpha\in\N^{r}}$ .

Then  the  pressure density  of the Multiscale SK model defined  in \eqref{multiscalep2} can be represented as

\be\label{multiscalep}
p_N(\beta)= \frac{1}{N}\,\E\,\log Z_{N}(\beta)
\ee

where
\be
Z_ {N}(\beta)\,=\,\sum_{(\sigma,\balpha)\in\Sigma_{N,r}}\, \nu_{\balpha}\,e^{\,H_N(\sigma,\balpha)}
\ee

Within this formalism the proof of the main theorem is obtained by means of the upper  bound  obtained by a suitable interpolation and a  lower  bound by the cavity method.
The main issue  is to understand  the distribution of $c_{12}$  under the limiting Gibbs measure. The key point is that  we have to know  the  joint probability distribution of the two covariances
$q_{N}$ and $\gamma$. This situation is very similar to the case of the Multispecies SK model \cite{BCMT, panmulti} where it turns out that the Hamiltonian  can be suitably perturbed in order to satisfy a  \textit{synchronization property} that allows to generate the joint probability of different overlaps functions using the same RPC.\\

\section{DYNAMICS}

In \cite{cuku2} a dynamical approach, in the Langevin, as well in the  Montecarlo setting, was proposed to study the \textit{multi-bath} equilibrium for a spin system. We notice that it's possible to modify the algorithm proposed there in order to get the equilibrium measure of the model considered here. Let start for simplicity with the case $r=1$ where the parameters of the model are  $\zeta_0$ and $\gamma_1$, the general case $r>1$ can be obtained trough a similar construction. The recursive definition \eqref{recZ} implies that the equilibrium measure can be written as

\be\label{eqm}
\mu_N(\sigma,J)= \mu_N(J)\mu_N(\sigma|J)
\ee

where
\be\label{eqm2}
\mu_N(\sigma|J)= \dfrac{e^{\,-\beta H_N(\sigma)}}{Z_N}
\ee

and
\be\label{eqm3}
\mu_N(J)=  \dfrac{(Z_{N})^{\zeta_0}}{\E (Z_{N})^{\zeta_0 }}
\ee

The algorithm given in \cite{cuku2} is modified as follows. Consider a system of $N$ spins $\sigma=(\sigma_i)_{i\leq N}$ with energy
\begin{equation}\label{energy}
H(\sigma,J)= \Gamma \sum_i J_{ij} \sigma_i \sigma_j  + \frac{k}{2} \sum_{ij} J_{ij}^2
\end{equation}
where $\Gamma, k>0$ are two real parameters and $J=(J_{ij})_{i,j\leq N}$ is a set of couplings. The dynamics goes as follows :
\begin{itemize}
\item the $\sigma_i$ evolve with any dynamics (Glauber, Monte Carlo) associated with energy \eqref{energy} and temperature $T=\frac 1 \beta$.
\item the $J_{ij}$ (for $i \le j$ and $J_{ij}=J_{ji}$) evolve according to the Langevin equation:

\begin{equation}\label{lang}
\gamma  \dot J_{ij} = -k  J_{ij}- \Gamma \sigma_i \sigma_j + \rho_{ij}(t)
\end{equation}

The  $\rho_{ij}$ are uncorrelated white noises with zero mean and variance $2T^*$ where $T^*=1/\beta^*$ represents the temperature of the second equilibrium bath.
\end{itemize}
Given the spin configuration $\sigma(t)$ then \eqref{lang} implies that
\begin{equation}
J_{ij}(t) =  \int dt' \; e^{-\frac{k}{\gamma}  (t-t')} (- \Gamma \sigma_i(t') \sigma_j(t')+ \rho_{ij}(t'))
\label{slow}
\end{equation}
Let us now define the timescale of the slow bath $\tau_o= \frac \gamma k$, and assume that the equilibration time of the dynamics of the spins $\sigma$
at given $J$  is bounded by $\tau_{eq}$. Clearly, $\tau_{eq}$ may depend on the size of the
system, and become infinite as $N \rightarrow \infty$.

Considering now the case of timescale separation $\tau_o \gg \tau_{eq}$ we may replace in (\ref{slow}) the product $\sigma_i(t') \sigma_j(t')$ with its average $\langle\sigma_i\sigma_j\rangle_J$ with respect to the stationary measure
%
\be\label{statm}
\mu(\sigma|J)= \dfrac{e^{-\beta\left( \Gamma \sum_{ij} J_{ij} \sigma_i \sigma_j\right)}}{Z(\beta,J)} \; .
\ee

Since
\begin{equation}
 \langle \sigma_i \sigma_j\rangle_J  =  -\frac{1}{\Gamma \beta}
\frac{\partial }{\partial J_{ij}} \log Z(\beta,J) \; ,
\end{equation}
the \eqref{lang} becomes:
\begin{equation}
\gamma  \dot J_{ij} = -k J_{ij}  + \frac{1}{ \beta}  \frac{\partial }{\partial J_{ij}}\log Z(\beta,J) + \rho_{ij}(t)
\label{heq1a}
\end{equation}
and provides the following stationary measure for the $J$:
\begin{equation}\label{eqm4}
\mu(J) = \frac{ e^{-\beta^*(\frac{1}{2} k \sum_{ij}  J_{ij}^2 - \frac{1}{ \beta}  \ln Z(\beta,{J}))}}{\tilde{Z}(\beta,\beta^*)}
\end{equation}
where $\tilde{Z}(\beta,\beta^*)$ is the normalization factor. It's easy to check that \eqref{eqm4} matches \eqref{eqm3} by setting $\beta^{*}/\beta=\zeta_0$ and $\beta^*k=1 / (\gamma_1)^2$.

A first remark is that if $T^*=T$ the system is in contact with an equilibrium bath of temperature $T$, whatever the timescales involved, or equivalently $ \beta^{*}/\beta=\zeta_0=1$ corresponds to the annealed regime where the variables $\sigma$ and $J$ thermalize together. Instead if $T^*$ goes to infinity then $\zeta_0$ goes to zero and the equilibrium is described by the usual quenched measure


\subsection{Stationarization}

What we have argued above is that if  the times $\tau_o$ are sufficiently long, and the system size $N$ is finite, the dynamic process above yields all the expectation values associated with the generating functional
$\frac{1}{\zeta_0}\log\E(Z^{\zeta_0})$. Let us discuss here, at a more phenomenological
level, what we expect to happen in the case that the $N \to \infty$ limit is taken first. When the $J_{ij}$ are kept fixed, the dynamics never becomes stationary (two time correlations $C(t, t_0)$ never become a function of $(t - t_0)$:
the system ages \cite{cuku4}. In order to understand what happens when the $J_{ij}$ continuously change, we need to recall the physical intuition we have of the SK model landscape. What we know about this landscape is that if we change the $J_{ij}$
by small amount $\dfrac{\sum_{ij}J_{ij}J'_{ij}}{\sqrt{\sum_{kl}J^2_{kl}\sum_{k'l'}J'^2_{k'l'}}}=\epsilon$

 where $\epsilon$ is a small number of $O(1)$, the structure of the lowest energy states
is completely reshuffled. This strongly suggests, and indeed there is numerical evidence for this, that if the $J_{ij}$ are
continuously changing at a certain rate, the dynamics enters a regime in which it becomes stationary $C(t, t') = C(t-t')$ (and yet, nonequilibrium), $E(t) = const$, and follows the evolution at that rate. For the SK model one  also believes, again based on the dynamic solution \cite{cuku4}, that when $\tau_0$  goes to infinity even after $N \to\infty$, the stationary value of the energy density is the same as the one with the limits reversed ( $N \to\infty $ after  $\tau_0\to \infty $). This commutation of limits need not hold for other models such as the $p$ − spin mean field model for $p > 2$, but,
importantly, is expected to happen for finite dimensional spin glasses with short range interactions, the Edwards-Anderson model, for example. The argument in the latter case is simple, and may be seen as a simple generalization of the one given in \cite{FMPP}  (more about this below): physical arguments indicate that metastable states of high free energy are unstable with respect to nucleation of a lower free-energy phase, the more unstable the higher they are. In a situation with slowly evolving couplings, there is a competition between the nucleation time $\tau_{nucleation}$
 (a decreasing function of the free energy difference with the ground state) , and the rate at which couplings change $\tau_0$ . One expects then, roughly speaking, that all states with $\tau_{nucleation}$ will be constantly relaxed, and that one
is left for $\tau_0\to\infty$ with states close in free energy to the equilibrium one.

\subsection {A phenomenological heat-exchange understanding}

We make now an even more severely phenomenological discussion, which we believe however is illuminating. The out
of equilibrium dynamics of the Sherrington Kirkpatrick model has been solved (to the level of rigour of physics) in the out of equilibrium regime \cite{cuku4}. The two main actors of this solution are the two-point correlation function $C(t, t')$ and the integrated response $\chi(t, t')$  to a magnetic field that has been acting during an interval $[t', t]$, i.e. the magnetization
at time $t$ per unit field acting on this interval. A result of the theory is that for large times, the parametric plot $\chi(t, t')$ versus $C(t, t')$
 tends to a function $\chi(C)$ that for the Sherrington Kirkpatrick model happens to be directly related to
the Parisi function $x(q)$ through
\begin{equation}
  \dfrac{d\chi}{dC}\Big|_{C=q}\,=\,x(q)
\end{equation}

Franz et al. \cite{FMPP} have argued that this coincidence between Parisi and dynamical relations must hold true for all finite dimensional systems. Note that this is very surprising, because the dynamical results are obtained in the $t, t'\to\infty$ limit taken after $N \to\infty$, while equilibrium results concern the opposite limit. Although these developments are for
the aging systems, the same can be done in the presence of multiple baths and the inversion of limits $\tau_0\to\infty$ and $ N \to \infty$.   Another development of dynamics is the fact (see \cite{cuku5}) that the function $x(q)$ has the interpretation of an effective
temperature that depends on the scale $q$, given by $T_{eff}(q) = \frac{T}{x(q)}$.

We are now in a position to interpret one of the main consequences of the solution of the model at the equilibrium presented in the previous section. Let us recall that any trial  parameter $x$ must satisfies the condition  $\lim_{c \to 0^{+}}x(c)=\zeta_0$, namely  the Parisi solution  starts at $x_{min} = \zeta_0$. It means that all effective temperatures $T_{eff} (q) = \frac{T}{x(q)}$
are
smaller or equal than the effective temperature of the multibath $T^*= 1/\beta^*= 1/\beta\zeta_0$. But this is just saying that the couplings $J_{ij}$ need to be ‘hotter’ than the internal effective temperatures of the system, something automatically guaranteed by the quenched case where $\beta^*=0$. This in turn means that the system does not give heat to the $J_{ij}$.\\

\subsection{Perspectives for simulations and bound in finite-dimensional systems.}

We want to emphasize that  if the arguments of stationarization and of commutation of the $N\to\infty $ and $\tau_0\to\infty$   limits hold, one is then in
a position to simulate a finite dimensional Edwards-Anderson model for any given $\zeta_0$, and check whether there is a transition line in the $T-\zeta_0$ plane, a necessary condition for arguing in favour of a Parisi RSB solution. One may would also be able to check the nature
of the transition, which in turn might give an indication on the nature of the low-temperature phase. Studying a glass problem around its transition is of course much more efficient that deep in the glass phase, as witnessed by the work done over the years for the deAlmeida-Thouless line.

\vskip 0.5 cm

%
%

{\bf Acknowledgements:} J.K. was  supported by  the Simons  Foundation Grant No 454943 P.C. and E.M. were partially supported by PRIN project N. 2015K7KK8L and Alma Idea Project 2018.

\end{document}